\documentclass[aps,prb,twocolumn,groupedaddress]{revtex4}
\usepackage{psfrag,graphicx}

\begin{document}
%\preprint{}

\title{Quantum Precursor of Shuttle Instability}



\author{D. Fedorets}
\email[]{dima@fy.chalmers.se}
\author{L.Y. Gorelik}
\author{R.I. Shekhter}
\author{M. Jonson}

%\thanks{}

\affiliation{Department of Applied Physics, Chalmers University of
  Technology \\ and G\"oteborg University, SE-412 96 G\"oteborg,
  Sweden}



\date{\today}
% -----------------------------------------------------------------------
\begin{abstract}
% -----------------------------------------------------------------------
The effects of a coupling between the quantized mechanical
vibrations of a quantum dot  and
coherent tunneling of electrons through a single level in the dot are
studied. The equation of motion for the reduced density operator
describing the vibrational degree of freedom is obtained.
 It is shown that  that the
expectation value of the displacement  is an oscillating function of
time with an exponentially increasing amplitude, which is the signature of
a quantum shuttle instability.
\end{abstract}

%73.23.HK % Coulomb blockade; single-electron tunneling
%73.40.GK % Tunneling
%85.35.Be % Quantum well devices (quantum dots, quantum wires, etc.)
%85.85.+j % Micro- and nano-electromechanical systems (MEMS/NEMS) and
         % devices
%\keywords{}

\maketitle

%\section{}
%\subsection{}
%\subsubsection{}

% If in two-column mode, this environment will change to single-column
% format so that long equations can be displayed. Use
%\begin{widetext}
%\end{widetext}

Nanoelectromechanical systems (NEMS), where  electronic and
 mechanical degrees of  freedom are coupled, have been attracting a
great deal of attention recently
\cite{Roukes00,Craighead00}.  An important example of such a
system is the nanoelectromechanical single-electron transistor
(NEM-SET) --- a structure where the movable conducting island is
elastically coupled to the electrodes (Fig.1).
\begin{figure}[htbp]
  \begin{center}
    \psfrag{x}{x}
    \psfrag{+eV}{$\mu_L=\frac{eV}{2}$}
    \psfrag{0}{$\mu_R=-\frac{eV}{2}$}
    \psfrag{Lead}{Lead}
    \psfrag{Dot}{Island}
    \includegraphics[width = 7cm]{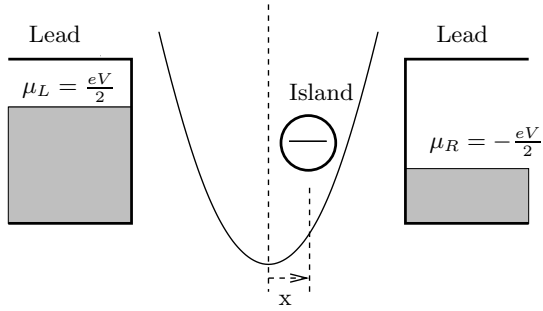}
    \vspace*{0.3cm}
    \caption{Model shuttle system consisting of a movable  conducting
    island placed
    between two leads.
    An effective elastic force acting on the dot from the leads is
    described by the parabolic potential.
      }\label{system}
  \end{center}
\end{figure}

In Ref.~\onlinecite{Gorelik98} it was shown,
that the NEM-SET becomes unstable
with respect to the development of periodic mechanical motion
 if a large enough bias voltage is applied between the
leads. This phenomenon is usually referred to as a shuttle
instability. The key issue in Ref.~\onlinecite{Gorelik98}
was that the charge on the
island, $q(t)$, is correlated with its velocity, $\dot{x}(t)$, in
such a way that the time average $ \overline{q(t)\dot{x}(t)} > 0$ even
if $ \overline{\dot{x}(t)} = 0$ (see the review
Ref.~\onlinecite{Jonson02}).
A classical theory of the
shuttle instability \cite{Gorelik98}
was based  on the assumptions that both
the charge on the island and
 its trajectory are well defined quantities.

A decrease of the island size should result in modifications of
the electromechanical phenomena in a NEM-SET as more quantum
mechanical effects come into play. There are two different types
of quantum effects which manifest itself as the island  size
decreases. The first one is the discreteness of the energy
spectrum. The electron energy level spacing in a nanometer-size
grain is of the order of $10$~K and resonant tunneling effects
become essential at small enough temperatures. In this case the
characteristic de Broglie wave length associated with the island
can still be much
shorter than the length scale of the spatial variations of the
"mechanical" potential.
If so, the motion of the island can be treated classically. The
NEM-SET in this regime has been studied theoretically in
Ref.~\onlinecite{Fedorets02} and the conditions for the shuttle
instability to appear have been found.

Diminishing the size of the island further results in the
quantization of the mechanical motion of the island. As a result
not only  the charge on the island but also its trajectory
experience strong quantum fluctuations and the picture of the shuttle
instability, which was elaborated in Refs.~\onlinecite{Gorelik98} and
\onlinecite{Fedorets02} is no longer valid.

A NEM-SET system in the regime of quantized mechanical
motion of the central island was studied theoretically in
Ref.~\onlinecite{Boese00}. It was assumed
that the phase breaking processes are strong enough to make the
density matrix  diagonal in the representation of the eigenstates
of the quantum oscillator Hamiltonian $|n\rangle $, which describes the
mechanical subsystem. At the same time it is well known that the
expectation value of the displacement operator in the eigenstates of a
quantum oscillator are zero while it is the coherent state, which
is a coherent superposition of all $|n\rangle $, that
approaches a description of the classical periodic motion of the
oscillator as $\hbar \rightarrow 0$.
 The interesting question arises, therefore,  how the
expectation value of displacement operator evolves in time and
what state
results when the formal condition of the
shuttle instability are satisfied for the quantum NEM-SET.

In this article we will show that if a large enough bias voltage is
applied between the leads, then the quantum state of the central
island of the NEM-SET evolves in such a way that the expectation value
of the displacement operator is not identically zero. Rather, it
oscillates in time with an increasing amplitude~\cite{Novotny02}.
This results
shows that the shuttle instability is a fundamental phenomenon
which exists even when the trajectory of the island and the charge
on it are no longer well-defined.

% --------------------------------------------------------------------

We use the following Hamiltonian to model our system
\begin{eqnarray}
H = \sum_{\alpha, k} \epsilon_{\alpha k} a_{\alpha k}^{\dag}
a_{\alpha k}+ \left[ \epsilon_0 -e {\cal E}X \right]c^{\dag} c +
\nonumber\\+ \frac{P^2}{2M} + \frac{Mw_0^2X^2}{2} + \sum_{\alpha, k}
T_{\alpha}(X) (a_{\alpha k}^{\dag} c + c^{\dag} a_{\alpha k})\,.
\end{eqnarray}
 The
first term in the Hamiltonian describes the electrons in the
electrodes, the second term relates to the single energy level
in the central
island, the third and forth terms to the quantized vibrational
degree of
freedom associated with center-of-mass motion of the central
island and the last term describes tunneling between the
electrodes and the island. All energies are measured from the Fermi
energy of the leads. Here we assume that only one single
electron state is available in the central island and that the
electrons in each electrode are non-interacting with a constant
density of states.

Let us introduce dimensionless operators for displacement,
$x\equiv X/r_0$,
and momentum, $p\equiv r_0 P/\hbar$, where
$r_0 \equiv \sqrt{\hbar/(Mw_0)}$, and then measure all lengths in
units of $r_0$ and all energies in units of $\hbar w_0$.

In order to to transfer the $x$-dependence of the island energy
level to the term describing tunneling,
we  make the unitary transformation
\begin{equation}\label{}
    \tilde{H} \equiv U H U^{\dag}\,,\quad U \equiv e^{i d
c^{\dag}c p}\,
\end{equation}
The Hamiltonian can now be written as
\begin{equation}\label{}
    \tilde{H} = H_e + H_v + \Omega\,,
\end{equation}
where
\begin{eqnarray}\label{}
    H_e &\equiv& \sum_{\alpha, k} \epsilon_{\alpha k} a_{\alpha k}^{\dag}
a_{\alpha k}+  \tilde{\epsilon}_0 c^{\dag} c \equiv H_a + H_c\,, \\
    H_v &\equiv& \frac{p^2}{2} + \frac{x^2}{2}\,,\\
    \Omega &\equiv& \sum_{\alpha, k} T_\alpha \left\{\Omega_{\alpha}^{\dag}
a_{\alpha k}^{\dag} c + h.c. \right\}\,,
\end{eqnarray}
with $\Omega_{\alpha} \equiv e^{idp} T_\alpha(x)\,$,
$\tilde{\epsilon}_0 \equiv \epsilon_0 - {Mw_0^2d^2}/{2}$ and
$d \equiv e{\cal E}/(Mw_0^2)$.

The evolution of the system is described by the
 Liouville-von Neumann equation for the total density operator
$\sigma$:
\begin{equation}\label{LN1}
    i  \partial_t \sigma =
[\tilde{H}, \sigma] = [H_e + H_v, \sigma] + [\Omega,
    \sigma]\,.
\end{equation}

It is convenient to transform Eq.~(\ref{LN1}) into the interaction
picture, defined by
\begin{equation}\label{}
    \tilde{\sigma}(t) \equiv e^{i(H_e+H_v) t}
{\sigma}(t) e^{-i(H_e+H_v) t}\,.
\end{equation}
Then
\begin{equation}\label{LN2}
    i\partial_t \tilde{\sigma}(t)  =  [\tilde{\Omega}(t),
    \tilde{\sigma}(t)]\,,
\end{equation}
where
\begin{equation}
\tilde{\Omega}(t) \equiv e^{i(H_e+H_v) t}
{\Omega}(t) e^{-i(H_e+H_v) t}\,.
\end{equation}

In what follows we will assume that  electrons in the leads are
weakly coupled to the rest of the system and that the leads are so
large that their statistical properties are unaffected by the weak
coupling. Then the following approximation can be made
\begin{equation}\label{approx1}
    \tilde{\sigma}(t) \approx \tilde{\rho}(t) \otimes \sigma_L \otimes
\sigma_R\,.
\end{equation}

Since we are interested in the evolution of the variables
describing
 the central island we need an equation of motion only for the reduced
 density operator. If we trace over the electrodes degrees of
 freedom directly in Eq.~(\ref{LN2}) we obtain zero in the RHS, which
 means that the effect is of the higher order with respect to $\Omega$.

If we formally integrate both sides of Eq.~(\ref{LN1}) and
substitute the result back into the  RHS of Eq.~(\ref{LN1}), we obtain
an integro-differential equation with RHS of the second order
with respect to $\Omega$,
\begin{equation}\label{LN3}
    \partial_t \tilde{\sigma}(t)  = -i [\tilde{\Omega}(t),
    \tilde{\sigma}(0)] -   \int_0^t dt_1\, [\tilde{\Omega}(t),
    [\tilde{\Omega}(t_1),
    \tilde{\sigma}(t_1)]]\,.
\end{equation}

We trace out the electrode degrees of freedom and get the
integro-differential equation for the reduced density operator
\begin{equation}\label{LN4}
    \partial_t \tilde{\rho}(t) =  -
{\rm Tr}_a \left\{\int_0^t dt_1\, \left[\tilde{\Omega}(t),
    \left[\tilde{\Omega}(t_1),
    \tilde{\sigma}(t_1)\right]\right]\right\}\,,
\end{equation}
where $\tilde{\rho}(t) \equiv {\rm Tr}_a \left\{ \tilde{\sigma}(t)
\right\}$ is the reduced density operator in the interaction
representation
 and the trace is over the
electrode degrees of freedom. The first term in Eq.~(\ref{LN3})
gives zero because we choose such initial condition that $
\sigma(0) = \rho(0) \otimes \sigma_L \otimes \sigma_R$.

By using  Eq.~(\ref{approx1})  we obtain
\begin{eqnarray}\label{}
    \partial_t \tilde{\rho}_t  =  -
  \sum_{\alpha}
    \int d\epsilon\, {\cal D}_\alpha
    \int_0^{t} dt_1\,
    e^{i\epsilon(t-t_1)} \nonumber\\
\times\left\{
    \left\{ (\tilde{\Omega}_{\alpha}^{\dag} \tilde{c})_t
    (\tilde{\Omega}_{\alpha} \tilde{c}^{\dag})_{t_1}
    \tilde{\rho}_{t_1}
     - (\tilde{\Omega}_{\alpha} \tilde{c}^{\dag})_{t_1}
\tilde{\rho}_{t_1} (\tilde{\Omega}_{\alpha}^{\dag}
\tilde{c})_t\right\}
f_{\alpha}^{+}   \right.\nonumber\\
\left.    +  \left\{ \tilde{\rho}_{t_1}(\tilde{\Omega}_{\alpha}
\tilde{c}^{\dag})_{t_1} (\tilde{\Omega}_{\alpha}^{\dag}
\tilde{c})_t - (\tilde{\Omega}_{\alpha}^{\dag} \tilde{c})_t
\tilde{\rho}_{t_1}
    (\tilde{\Omega}_{\alpha} \tilde{c}^{\dag})_{t_1}\right\}
f_{\alpha}^{-} \right\}\nonumber\\ + h.c.\,,
\end{eqnarray}
where $f_{\alpha}^{+} \equiv f_{\alpha} \equiv [1+e^{\beta
  (\epsilon-\mu_\alpha)}]^{-1}$, $f_{\alpha}^{-}
\equiv 1 - f_{\alpha}$ and ${\cal D}_\alpha$ is the density of
states in the corresponding lead.

This integro-differential equation becomes a differential equation
if we consider the case of zero temperature, $T=0$, and large bias
voltage, $eV \rightarrow \infty$,
\begin{eqnarray}\label{}
\partial_t \tilde{\rho}_t =  \pi {\cal D}_L
      \left[2(\tilde{\Omega}_{L} \tilde{c}^{\dag})_{t} \tilde{\rho}_{t}
      (\tilde{\Omega}_{L}^{\dag} \tilde{c})_t -
\left\{ (\tilde{\Omega}_{L}^{\dag} \tilde{c})_t
    (\tilde{\Omega}_{L} \tilde{c}^{\dag})_{t}, \tilde{\rho}_{t} \right\}
       \right]
      \nonumber \\
     + \pi {\cal D}_R \left[ 2(\tilde{\Omega}_{R}^{\dag}
\tilde{c})_t \tilde{\rho}_{t} (\tilde{\Omega}_{R}
\tilde{c}^{\dag})_{t}- \left\{(\tilde{\Omega}_{R}
\tilde{c}^{\dag})_{t} (\tilde{\Omega}_{R}^{\dag} \tilde{c})_t ,
\tilde{\rho}_{t}\right\} \right]\,.
\end{eqnarray}

To describe the evolution of the oscillator variables we need only
$\tilde{\rho}_{0,0} \equiv <0|\tilde{\rho}|0>$ and
$\tilde{\rho}_{1,1} \equiv <1|\tilde{\rho}|1>$, where $|1> \equiv
c^{\dag} |0>$. The corresponding equations for
 $\tilde{\rho}_{0,0}$ and $\tilde{\rho}_{1,1}$ are given by
\begin{eqnarray}\label{}
 \partial_t \tilde{\rho}_{0,0} &=& - \pi {\cal D}_L
      \left\{ \tilde{\Omega}_{L}^{\dag}
    \tilde{\Omega}_{L} , \tilde{\rho}_{0,0} \right\}
     + 2\pi {\cal D}_R\, \tilde{\Omega}_{R}^{\dag}  \tilde{\rho}_{1,1}
    \tilde{\Omega}_{R} \,, \\
 \partial_t \tilde{\rho}_{1,1}  &=&  -  \pi {\cal D}_L
\left\{\tilde{\Omega}_{R}
 \tilde{\Omega}_{R}^{\dag},
\tilde{\rho}_{1,1}\right\} + 2\pi {\cal D}_L\, \tilde{\Omega}_{L}
\tilde{\rho}_{0,0} \tilde{\Omega}_{L}^{\dag}\,.
\end{eqnarray}

It is convenient to change from  $\tilde{\rho}_{0,0}$ and
$\tilde{\rho}_{1,1}$ to $R_{0}$ and $R_{1}$ given by
\begin{eqnarray}\label{}
    R_{0}(t) &\equiv& e^{i\frac{d}{2}p}
     e^{-iH_v t} \tilde{\rho}_{0,0}e^{iH_v t}
    e^{-i\frac{d}{2}p}\,,\\
    R_{1}(t) &\equiv& e^{-i\frac{d}{2}p}
    e^{-iH_v t} \tilde{\rho}_{1,1}e^{iH_v t}
    e^{i\frac{d}{2}p}\,.
\end{eqnarray}
Then
\begin{eqnarray}
\partial_t R_{0} =
-i\left[H_v\left(x+\frac{d}{2}\right),R_{0}\right]
-\frac{1}{2}
      \left\{ \tilde{\Gamma}_L(x), R_{0} \right\} \nonumber\\
     + \sqrt{\tilde{\Gamma}_R(x)}\, R_{1} \sqrt{\tilde{\Gamma}_R(x)}
\,,\label{R:0}\\
 \partial_t R_{1}  =
 -i\left[H_v\left(x-\frac{d}{2}\right),R_{1}\right] -
\frac{1}{2}
  \left\{\tilde{\Gamma}_R(x),
R_{1}\right\} \nonumber\\ + \sqrt{\tilde{\Gamma}_L(x)}\,
R_{0}\sqrt{\tilde{\Gamma}_L(x)}\,,\label{R:1}
\end{eqnarray}
where $\tilde{\Gamma}_{L, R}(x) \equiv \Gamma_{L, R}(x+d/2)$ and
$\Gamma_{\alpha}(x) \equiv 2\pi {\cal
  D}_\alpha T_\alpha^2(x)$.
 These two equations completely describe
the evolution of the vibrational degree of freedom.

By using Eqs.~(\ref{R:0})  and (\ref{R:1}) we can obtain the equations of
motion for any momenta
with respect to the density operators $R_{+}\equiv R_0 + R_1$  and
$R_{-}\equiv R_0 - R_1$.
We expand $\tilde{\Gamma}_{L,R}(x)$  to first
order with respect to the displacement,
$\tilde{\Gamma}_{L,R}(x) = (1\mp x/\lambda)\tilde{\Gamma}_{L,R}(0)$, where
$\lambda$ is characteristic tunneling length, and leave only terms of
the first order with respect to $\lambda^{-1}$.

In the quasisymmetric case,
 $\tilde{\Gamma}_L(0) = \tilde{\Gamma}_R(0)$, the equations for
 $n_{-}$, $x_{+}$ and $p_{+}$ are decoupled from the rest,
\begin{eqnarray}
 \dot{x}_{+} &=& p_{+}\,,\label{x:plus}\\
\dot{p}_{+} &=& - x_{+} - d\, n_{-}\,,\label{p:plus}\\
 \dot{n}_{-} &=&   -
\tilde{\Gamma} n_{-} + \frac{2\tilde{\Gamma}}{\lambda}x_{+}\,
\label{n:minus}
\end{eqnarray}
and equations for the second momenta are decoupled from the higher
momenta,
\begin{eqnarray}
  \partial_t\left<x^2\right>_{+} &=&
\left<\left\{p,x\right\}\right>_{+}\,,\label{x^2:plus}\\
\partial_t\left<p^2\right>_{+} &=& - \left<\left\{p,x
  \right\}\right>_{+} - 2d p_{-}\,,\label{p^2:plus}\\
 \partial_t\left<\left\{p,x\right\}\right>_{+}
&=&  - 2\left<x^2\right>_{+} +
 2\left<p^2\right>_{+} - 2dx_{-}
\,,\label{px:plus}\\
\dot{x}_{-} &=&  p_{-} - \tilde{\Gamma} x_{-} +
 \frac{2\tilde{\Gamma}}{\lambda} \left<x^2 \right>_{+}\,, \label{x:minus}\\
\dot{p}_{-} &=& - x_{-} -  d  -
\tilde{\Gamma} p_{-} + \frac{
\tilde{\Gamma}}{\lambda}\left<\left\{p,x\right\}\right>_{+}\,, \label{p:minus}
\end{eqnarray}
where $\left< \bullet\right>_{\pm} \equiv {\rm Tr}\{R_{\pm}\bullet\}$,
 $n_{-}\equiv \left<1\right>_{-}$,
$x_{\pm} \equiv \left<x\right>_{\pm}$ and $p_{\pm} \equiv
\left<p\right>_{\pm}$ and
$\tilde{\Gamma} \equiv \tilde{\Gamma}_L(0) + \tilde{\Gamma}_R(0)$.

The characteristic equation for the system of Eqs.~(\ref{x:plus},
\ref{p:plus},\ref{n:minus}) is given by
\begin{eqnarray}
 (\alpha^2+1)(\alpha+\tilde{\Gamma})+2\gamma = 0\,,\quad
\gamma \equiv \frac{d}{\lambda}\tilde{\Gamma} \ll 1\,
\end{eqnarray}
and has three roots
\begin{eqnarray}
  \alpha_1 &\approx& -\tilde{\Gamma}
\left[1+\frac{2\gamma}{\tilde{\Gamma}^2+1}\right]\,, \\
  \alpha_2 &\approx& i + \frac{\gamma}{1-i\tilde{\Gamma}}\,,\,
  \alpha_3 \approx \alpha_2^{*}\,.
\end{eqnarray}
The first root is a negative real number, which corresponds to a solution
which exponentially goes to zero.
The last two roots have positive real parts and non-zero imaginary
parts, which gives rise to oscillating solutions with exponentially
increasing amplitudes.
The expectation value $\overline{x}(t)\equiv {\rm Tr}
\left\{ U^{\dag}\sigma(t)U x\right\}$
 of the displacement operator $x$ depends on $x_{+}$
as follows: $\overline{x}(t) = x_{+} + {d}/{2}$. This means that when
we apply a high enough bias voltage between the leads the
expectation value of the displacement starts to oscillate with an
increasing amplitude with respect to the point $x=d/2$ of the original
coordinate system.

The characteristic equation for the system of Eqs.~(\ref{x^2:plus},
\ref{p^2:plus},\ref{px:plus},\ref{x:minus},\ref{p:minus}) is given by
\begin{equation}\label{}
    \alpha[4+\alpha^2][1+(\tilde{\Gamma}+\alpha)^2] +
2\gamma[4\tilde{\Gamma} \alpha+5\alpha^2-4]=0\,,
\end{equation}
and has five roots
\begin{eqnarray}
  \alpha_1 &\approx&  \frac{2\gamma}{\tilde{\Gamma}^2+1} \,, \\
  \alpha_2 &\approx& 2i + \frac{2\gamma}{1-i\tilde{\Gamma}}\,,\,
   \alpha_3 \approx \alpha_2^{*}\,, \\
\alpha_4 &\approx&  -\tilde{\Gamma} + i\,,\,\alpha_5 \approx \alpha_4^{*}\,.
\end{eqnarray}
The first root gives a non-oscillatory solution of exponentially growing
amplitude. The remaining four roots correspond to  oscillatory solutions with
increasing ($\alpha_2$ and $\alpha_3$) and decreasing amplitudes
($\alpha_4$ and $\alpha_5$), respectively. Thus the energy of the oscillator, which
is given by the sum of the two second momenta $\left<x^2\right>_{+}$
and  $\left<p^2\right>_{+}$, exponentially grows with time.

The importance of the position dependence of $T_\alpha(x)$ can be
seen if we let $\lambda \rightarrow \infty$ in the above treatment.
Then we find that the energy grows linearly with time, while the
average displacement does not grow.

In conclusion, we have studied a quantum shuttle system where
the quantized mechanical
vibrations of a quantum dot  is coupled to
coherent tunneling of electrons through a single level in the dot.
We have obtained the equation of motion for the reduced density operator
describing the vibrational degree of freedom.
It was shown
that the
expectation value of the displacement  is an oscillating function of
time with an exponentially increasing amplitude, which is the signature of
a quantum shuttle instability.

% --------------------------------------------------------------------

\end{document}